\documentclass[structabstract]{aa}
\usepackage{natbib}
\usepackage{txfonts}
\usepackage{graphicx}

\begin{document}

\title{Using limb darkening to measure fundamental parameters of stars}
\titlerunning{Limb darkening and fundamental parameters.}
\authorrunning{H.~R. Neilson \& J.~B. Lester}

\author{Hilding R. Neilson\inst{1} \and John B. Lester \inst{2,3}}
\institute{
     Argelander Institute for Astronomy, University of Bonn \\
     \email{hneilson@astro.uni-bonn.de}
 \and 
     Department of Chemical and Physical Sciences, 
     University of Toronto Mississauga 
 \and
     Department of Astronomy \& Astrophysics, University of Toronto \\
     \email{lester@astro.utoronto.ca}
}

\date{}

\begin{abstract}
{Limb darkening is an important tool for understanding stellar 
 atmospheres, but most observations measuring limb darkening 
 assume various parameterizations that yield no significant 
 information about the structure of stellar atmospheres.}
{We use a specific limb-darkening relation to study how the best-fit 
 coefficients relate to fundamental stellar parameters from spherically 
 symmetric model stellar atmospheres.}
{Using a grid of spherically symmetric \textsc{Atlas} model atmospheres,
 we compute limb-darkening coefficients, and develop a novel 
 method to  predict fundamental stellar parameters. 
 }
{We find our proposed method predicts the mass of stellar atmosphere models given only the radius and limb-darkening coefficients, suggesting that microlensing, interferometric, transit and eclipse observations can constrain stellar masses.}
{This novel method demonstrates that limb-darkening parameterizations 
 contain important information about the structure of stellar 
 atmospheres, with the potential to be a valuable tool for measuring 
 stellar masses.}
\end{abstract}

\maketitle

\section{Introduction} \label{sec:intro}

Limb darkening, the change of surface intensity from the center to the 
edge of a stellar disk, is a powerful measure of the physical structure 
of stellar atmospheres. However, it is difficult to observe the 
intensity profile across a stellar disk except for the Sun.  Most 
observations of stellar limb darkening are indirect, coming from light 
curves of eclipsing binaries \citep{Claret2008} or planetary transits 
\citep{Knutson2007}, interferometric visibilities \citep{Haubois2009} 
and microlensing light curves \citep{An2002}, and these indirect 
methods have limited precision \citep[e.g.][]{Popper1984, Zub2011}.  
Because of this, limb darkening is commonly treated as a parameterized 
function of $\mu \equiv \cos \theta$, where $\theta$ is the angle 
between the direction to the distant observer and the normal direction
at each location on the stellar surface.  An example of a linear 
parameterization is \citep{Schwarzschild1906}
\begin{equation}
\frac{I_\lambda(\mu)}{I_\lambda(\mu=1)} = 1 - a_\lambda(1-\mu).
\end{equation}

As numerical model atmospheres became robust, the computed intensity 
profiles were represented by more elaborate limb-darkening 
parameterizations that include higher order terms of $\mu$ or are 
expressed in terms of powers of $r = \sin \theta$ \citep{Heyrovsky2007}, as well as being 
normalized with respect to the stellar flux instead of the central 
intensity \citep{Wade1985}.  These more detailed limb-darkening laws were used to 
interpret the improving observations noted above.  

However, even with these advances, current limb-darkening observations 
are still unable to constrain the model stellar atmospheres. Recent 
interferometric observations of nearby red giants are not yet precise 
enough to differentiate between the center-to-limb intensity profiles 
from \textsc{Phoenix} and \textsc{Atlas} model atmospheres or even 
between plane-parallel or spherically symmetric models 
\citep{Wittkowski2004, Wittkowski2006b, Wittkowski2006a, Neilson2008}.  
However, the combination of the observations and models \emph{do} 
provide constraints for fundamental stellar parameters such as the 
effective temperature and gravity.

In this work, we continue our previous analysis 
\citep[][hereafter Paper 1]{Neilson2011} to explore the connection 
between limb darkening and stellar parameters.  We adopt the 
flux-conserving law 
\begin{equation} \label{eq:ldlaw}
\frac{I_\lambda(\mu)}{2\mathcal{H}_\lambda} = 1 - 
  A_\lambda \left (1 - \frac{3}{2}\mu \right ) -
  B_\lambda \left (1 - \frac{5}{4} \sqrt{\mu} \right ),
\end{equation} 
where $\mathcal{H}_\lambda$ is the Eddington flux defined as
\begin{equation}
\mathcal{H}_\lambda \equiv \frac{1}{2} \int_{-1}^{1} I_\lambda(\mu) \mu \mathrm{d} \mu,
\end{equation}
because this law was used by \citet{Fields2003} to analyze the 
microlensing event EROS-BLG-2000-5 \citep{An2002}.  We fit this 
limb-darkening law to the intensities of spherically symmetric model 
stellar atmospheres computed with the \textsc{SAtlas} code 
\citep{Lester2008} to explore the relation between fundamental parameters and limb-darkening coefficients hinted at in Paper 1.

\section{Limb darkening and fundamental parameters}
\label{sec:fundpar}

Many previous studies have used the limb darkening predicted by model 
stellar atmospheres as a tool to achieve a better determination of a 
stellar diameter \citep{Merand2010} or a better characterization of a 
transiting planet \citep{Lee2012}.  Our goal is different; we want to 
determine how limb darkening, represented by Eq.~(\ref{eq:ldlaw}), 
depends on the fundamental parameters of stellar atmospheres.  Because 
there are several steps leading to our conclusion, we describe our 
approach step-by-step to make it as clear as possible.

\subsection{Intensity fixed point}
\label{subsec:int_fp}

\begin{figure}[t]
\begin{center}
  \resizebox{\hsize}{!}{\includegraphics{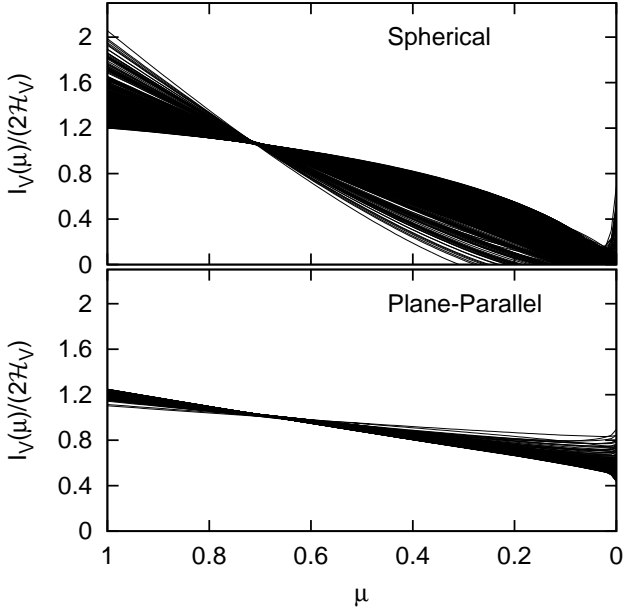}
                   }
  \caption{Best-fit $V$-band limb-darkening relations using 
           Eq.~(\ref{eq:ldlaw}) for grids of $5~M_{\sun}$ spherically symmetric 
           and plane-parallel \textsc{Atlas} model atmospheres. A fixed 
           $\mu$-point is clearly seen in both cases.}
  \label{fig:v-band-imu}
\end{center}
\end{figure}

Figure~\ref{fig:v-band-imu} plots the curves produced by the 
limb-darkening law given by Eq.~(\ref{eq:ldlaw}) for the $V$-band 
intensities computed for both plane-parallel and spherical 
\textsc{Atlas} models.  The cube of spherical models is from Paper 1, 
with luminosities and radii corresponding to the range of 
$3000~\mathrm{K} \leq T_\mathrm{eff} \leq 8000~\mathrm{K}$ in steps of 
$100~\mathrm{K}$, $-1 \leq \log g \leq 3$ in steps of $0.25$ and 
$2.5 \leq M/M_{\sun} \leq 10$ in steps of $2.5~M_{\sun}$, which we have 
supplemented with additional models having masses of $M = 0.5$ and 
$1.0~M_{\sun}$.  The grid of plane-parallel models spans the equivalent 
range of effective temperature and gravity.
Figure~\ref{fig:v-band-imu} shows that both sets of models have a 
fixed point, $\mu_1$, where the curves intersect, although there is 
a larger spread for the spherical models because of the combination of 
the three fundamental stellar parameters, $L_\star$, $M_\star$ and 
$R_\star$.  The immediate vicinity of the spherical fixed point is 
displayed in more detail in Fig.~6 of Paper 1.

\subsection{Relationship between $A$ and $B$}
\label{subsec:ab}

In Paper 1 we showed that the fixed point, $\mu_1$, is caused by the 
limb-darkening law's $A_\lambda$-coefficient being linearly related to the 
$B_\lambda$-coefficient, $A_\lambda \propto \alpha_\lambda B_\lambda$.  The proportionality term,
$\alpha_\lambda$, depends \emph{only} on two quantities, both of which are functions 
of the stellar intensity: the mean intensity, $J_\lambda$, and the 
pseudo-moment $\mathcal{P}_\lambda$, defined as 
\begin{equation} \label{eq:def_p}
\mathcal{P}_\lambda \equiv \frac{1}{2\mathcal{H}_\lambda}\int I_\lambda\sqrt{\mu}d\mu.
\end{equation}
Combining these two radiation terms defines a new quantity,
\begin{equation} \label{eq:def_eta}
\eta_\lambda \equiv \frac{\mathcal{P}_\lambda}{J_\lambda},
\end{equation}
which is similar to the Eddington factor, $f = K/J$ \citep{Mihalas1978}.
As we showed in Eq.~(19) of Paper 1, $\alpha_\lambda$ 
depends only on $\eta_\lambda$,
\begin{equation} \label{eq:alpha}
\alpha_\lambda = -\frac{(5\eta_\lambda/4 - 1)/6 + 11/96}
                       {(5\eta_\lambda/4 - 1)/4 + 1/6}.
\end{equation}

\subsection{Relationship between $\alpha$ and $\mu_1$}
\label{subsec:alpha_mu1}

Our analysis in Paper 1 also showed that the fixed point, $\mu_1$, is 
a function of $\alpha_\lambda$.  Rearranging our result from Paper 1, 
we obtain 
\begin{equation} \label{eq:mu1}
\mu_1 = \frac{3\alpha_\lambda(1+\alpha_\lambda) + \frac{25}{16} + 
              \sqrt{\frac{75}{8}\alpha_\lambda(1+\alpha_\lambda) +
              \frac{625}{256}}}
             {\frac{9}{2} \alpha_\lambda^2}.
\end{equation}
Working back through this chain of logic, we find that because $\mu_1$ 
depends on $\alpha_\lambda$, which depends on $\eta_\lambda$, we are 
led to the conclusion that $I_\lambda(\mu_1)/2\mathcal{H}_\lambda$ 
depends solely on the two angular moments of the stellar intensity, 
$J_\lambda$ and $\mathcal{P}_\lambda$.

 The analytic derivations of Eq.~(\ref{eq:alpha}) and 
Eq.~(\ref{eq:mu1}) in Paper~1 assumed, for convenience, that 
$\eta_\lambda$ is the same for all models.  This assumption is true for 
plane-parallel models because they share the same flat geometry. 
However, $\eta_\lambda$ varies for spherical atmospheres because they 
have different amounts of curvature depending on the values of their 
particular fundamental parameters. As a result, the values of 
$\alpha_\lambda$ and $\mu_1$ also vary for spherical atmospheres.  
However, this variation does not alter the logical connections between 
these quantities, as we show numerically in the following section. 

\subsection{Dependence on surface gravity}
\label{subsec:fgrav}

\begin{figure*}[t]
  \centering
  \resizebox{\hsize}{!}{\includegraphics{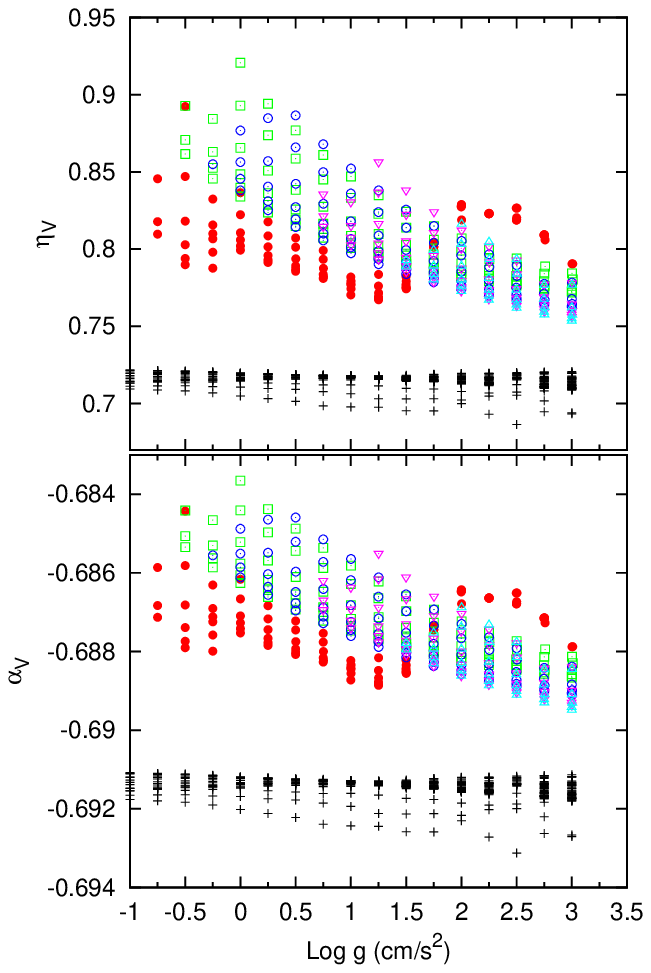}
                        \includegraphics{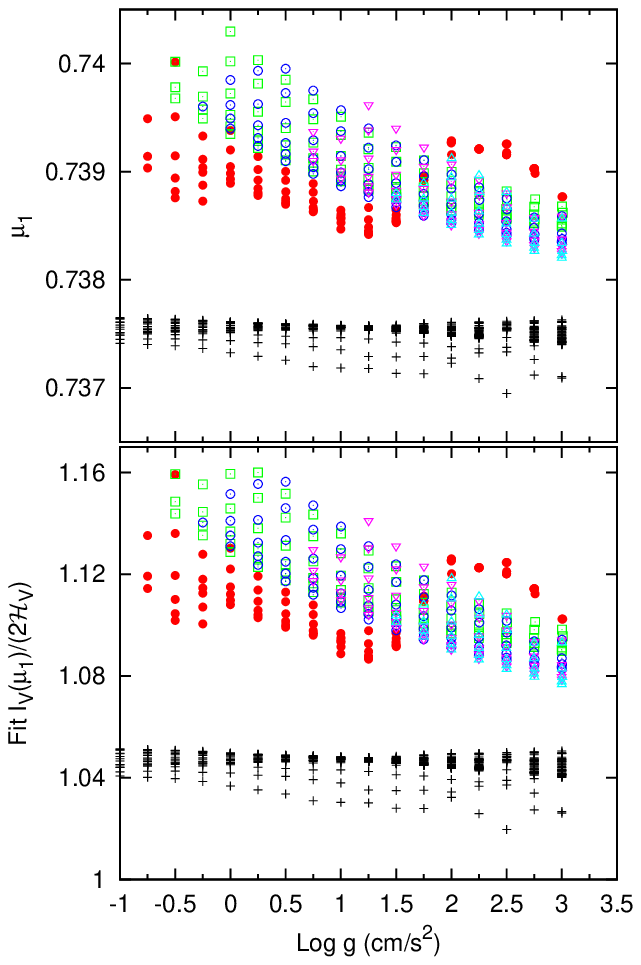}}
  \caption{Surface gravity dependence of $\eta_\lambda$, 
           $\alpha_\lambda$, fixed point $\mu_1$ and the normalized 
           intensity at the fixed point,
           $I_\lambda(\mu_1)/2\mathcal{H}_\lambda$, for the $V$ 
           spectral band.  Plane-parallel model 
           atmospheres are denoted by black crosses, and spherical 
           model atmospheres are represented by other symbols.  Filled 
           red 
           circles are $T_\mathrm{eff} = 3000$~K, open green squares 
           4000~K, open blue circles 5000~K, downward pointing magenta 
           triangles 6000~K and upward pointing pale blue triangles are 
           7000~K.}
  \label{fig:v-band-logg}
\end{figure*}

Having used our analytic results to trace the logical 
connection between the key variables, we now continue our exploration 
using large grids of plane-parallel and spherical model atmospheres to 
compute the key quantities numerically, which enables us to drop the 
assumption that $\eta_\lambda$ is constant.  
Figure~\ref{fig:v-band-logg} shows the results for the $V$-band.

Beginning with the plane-parallel models, the black crosses in the 
upper left plot of Fig.~\ref{fig:v-band-logg} shows that the $\eta$ 
parameter is nearly constant over a very wide range of surface gravity 
for the models being considered.  As required by Eq.~(\ref{eq:alpha}), 
the near constancy of $\eta$ causes the $\alpha$ coefficient also to be 
almost constant, as shown by the black crosses in lower left panel of 
Fig.~\ref{fig:v-band-logg}.  Next, as a result of Eq.~(\ref{eq:mu1}), 
the fixed point, $\mu_1$, must be nearly constant, as shown by the 
black crosses in the upper right panel of Fig.~\ref{fig:v-band-logg}.  
Finally, when the nearly constant $\mu_1$ is used in 
Eq.~(\ref{eq:ldlaw}), the result is a nearly constant value of the 
normalized intensity at the fixed point.  This conclusion is confirmed 
by the black crosses in the lower right plot of 
Fig.~\ref{fig:v-band-logg}.

The spherically symmetric models are fundamentally different.  The 
various colored symbols in the upper left plot in 
Fig.~\ref{fig:v-band-logg} confirms the result of 
Sect.~\ref{subsec:alpha_mu1} that the parameter $\eta$ \emph{does} 
vary with surface gravity.  It then follows from Eq.~(\ref{eq:alpha}) 
that $\alpha$ must also vary, as shown by the colored symbols in the 
lower left plot.  As before, Eq.~(\ref{eq:mu1}) requires that the 
variation of $\alpha$ leads to the variation of $\mu_1$ shown by the 
colored symbols in the upper right plot of Fig.~\ref{fig:v-band-logg}.
Finally, this leads to the variation of the normalized intensity at the 
fixed point shown by the colored symbols in the lower right plot.
For spherically symmetric models, the conclusion is that the location 
and intensity of the fixed point depend on the fundamental stellar 
parameters.

The beginning step of this logical chain, the dependence of 
$\eta$ on the atmospheric parameters, is physically reasonable.  As shown in the upper
left panel of Fig.~\ref{fig:v-band-logg}, as $\log g$ decreases,
$\eta$ increases towards unity, meaning that 
$\mathcal{P} \rightarrow J$.  This is the result of the intensity profile becoming 
extremely centrally concentrated, or, equivalently, the 
atmospheric extension approaches the 
stellar radius. Because $\eta$ depends upon the extension of the 
atmosphere, so do $\alpha$, the fixed point $\mu_1$, and the intensity 
at the fixed point.

\section{Parameterization of atmospheric extension}
\label{sec:atm_ext}

\begin{figure}[t!]
  \centering
  \resizebox{\hsize}{!}{\includegraphics{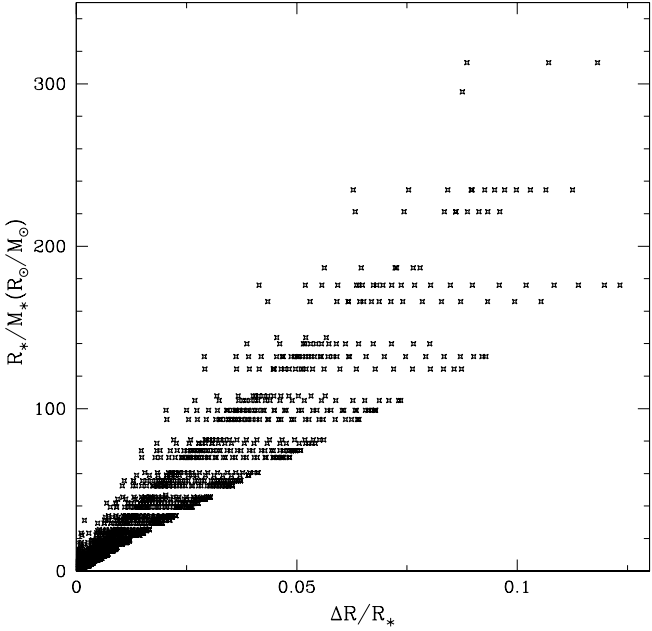}}
  \caption{Extension parameter $R_\star/M_\star$ compared to the 
           relative atmospheric extension, $\Delta R/R_\star$, 
           derived for more than 3000 spherical models.}
  \label{fig:extension}
\end{figure}
To explore further the results shown in Fig.~\ref{fig:v-band-logg}, we 
replace $\log g$ by a more explicit representation of the 
extension of the stellar atmosphere, for which we adopt 
\begin{equation} \label{eq:exten}
\frac{\Delta R}{R_\star} \propto \frac{R_\star}{M_\star}.
\end{equation}
This dependence is physically plausible because the extension is larger 
for stars of larger radius, but greater masses pull the atmospheres 
into a more compact configuration.

One way of deriving Eq.~(\ref{eq:exten}) begins 
with the definition of the pressure scale height,
\begin{equation}
H \equiv \frac{k T}{\mu m_\mathrm{H} g},
\end{equation}
where $H$ is the distance over which the pressure changes by a factor 
of $e$, $k$ is the Boltzmann constant, $T$ is the gas temperature, 
$\mu$ is the mean molecular weight of the gas, $m_\mathrm{H}$ is the 
mass of the hydrogen atom, and the surface gravity 
$g = G M_\star/R^2_\star$. Using $H$ to represent the thickness of the 
atmosphere, the relative extension of the atmosphere is 
\begin{equation}
\frac{\Delta R}{R_\star} = \frac{H}{R_\star},
\end{equation}
both sides of which are dimensionless.  Assuming that $T$ and $\mu$ are 
nearly constant over the distance $H$, the relative extension of the 
atmosphere becomes 
\begin{equation}
\frac{\Delta R}{R_\star} 
         \approx \frac{\rm{constant}}{R_\star M_\star/R^2_\star}
         \propto \frac{R_\star}{M_\star},
\end{equation}
which is our expression.

To test the correlation of $R_\star/M_\star$  with the actual 
atmospheric extension, we use the Rosseland optical depth to determine 
for each of our several thousand spherical models the quantity 
$\Delta R/R_\star$, where $\Delta R$ is the physical distance from the 
location where $\tau_\mathrm{Ross} = 1$ out to the location of where 
$\tau_\mathrm{Ross} = 0.001$, which is slightly more than one 
pressure scale height, and $R_\star$ is the radius where 
$\tau_\mathrm{Ross} = 2/3$.  Figure~\ref{fig:extension} plots the 
extension parameter $R_\star/M_\star$ for each model versus that 
model's $\Delta R/R_\star$.  While there is a spread because $T$ and 
$\mu$ do vary over the distance $H$, it is obvious that 
$R_\star/M_\star$ tracks the actual extension of the models quite well. 

\begin{figure*}[t]
  \centering
  \resizebox{\hsize}{!}{\includegraphics{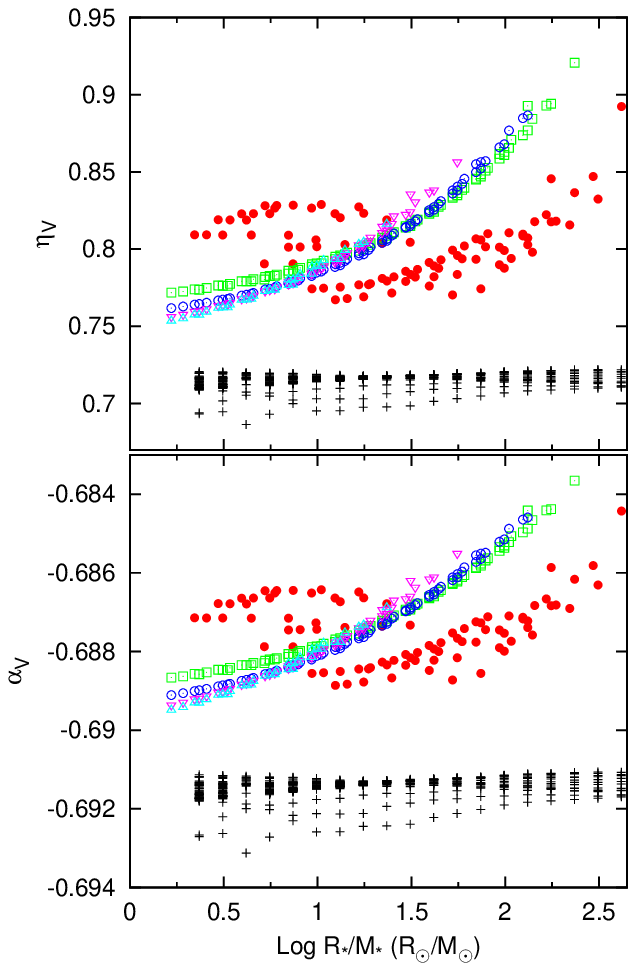}
                        \includegraphics{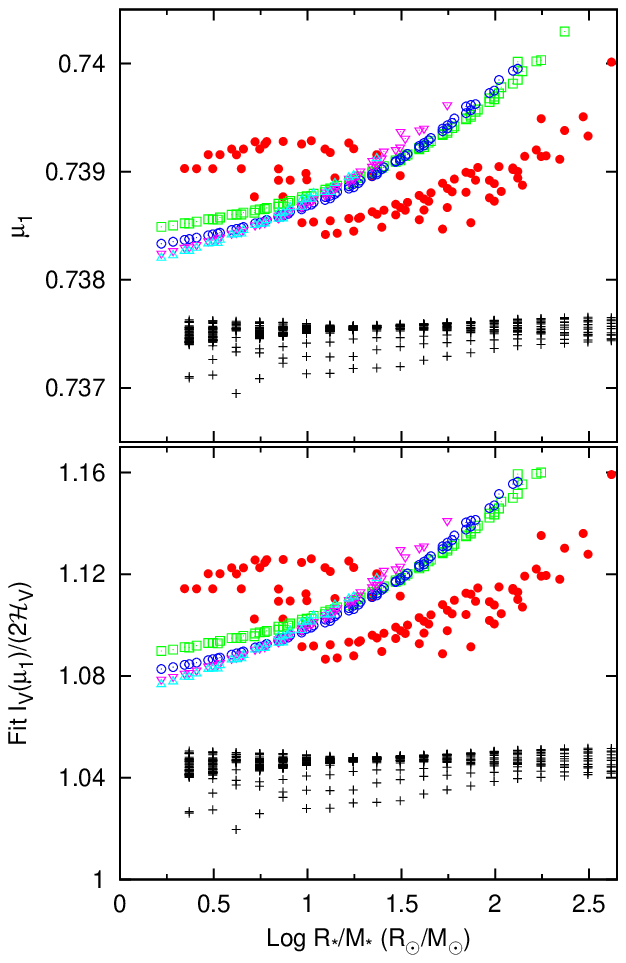}}
  \caption{(Left) Values of the pseudo-moment $\eta_\mathrm{V}$ and variable 
           $\alpha_\mathrm{V}$ as a function of atmospheric extension for spherical
           atmospheres, $R_\star/M_\star$ in solar units.
           (Right) The top panel shows the value of the primary fixed 
           point, $\mu_1$, for the linear-plus-square-root 
           parametrization, also plotted as a function of the 
           extension of the spherical atmosphere. 
           The bottom panel shows the dependence of the normalized 
           $V$-band intensity of the fixed point on the atmospheric 
           extension. The symbols have the same meaning as in 
           Fig.~\ref{fig:v-band-logg}. When comparing to 
           Fig.~\ref{fig:v-band-logg}, note that increasing extension 
           corresponds to decreasing $\log g$.}
  \label{fig:fp-ext}
\end{figure*}

Using $R_\star/M_\star$ to represent the atmospheric extension 
transforms Fig.~\ref{fig:v-band-logg} into Fig.~\ref{fig:fp-ext}, 
where the extension parameter for the plane-parallel models is 
determined by assuming a mass of $5~M_{\sun}$ and computing the radius 
for each model from its surface gravity.  
Figure~\ref{fig:fp-ext} shows, as expected, that $\eta_\lambda$ 
is a tight function of $\log (R_\star/M_\star)$, except for the models 
with $T_\mathrm{eff} = 3000$~K.  It follows, because of the connections 
established in Sec.~\ref{sec:fundpar}, that all other quantities also 
depend on $R_\star/M_\star$.  For plane-parallel atmospheres the 
quantities are nearly constant, as expected, and as they were found to 
be in Fig.~\ref{fig:v-band-logg}.
Therefore, the values of $\mu_1$ and 
$I_\lambda(\mu_1)/2\mathcal{H}_\lambda$ depend on the fundamental 
stellar parameters 
$R_\star$ and $M_\star$.  The deviation of the models with 
$T_\mathrm{eff} = 3000$~K may be due to changes in the atmospheres 
at the lowest temperature.

Excluding the models with $T_\mathrm{eff} = 3000~\mathrm{K}$, we fit 
for each waveband the dependence of $\mu_1$ and $I_\lambda(\mu_1)/2\mathcal{H}_\lambda$ on
$\log (R_\star/M_\star)$ shown in Fig.~\ref{fig:fp-ext} using the 
functional forms 
\begin{equation} \label{eq:mu1_rm}
\mu_1 = C_\mu \left(\log \frac{R_\star}{M_\star} \right )^2 + D_\mu,
\end{equation}
and
\begin{equation}\label{eq:imu}
\frac{I_\lambda(\mu_1)}{2\mathcal{H}_\lambda} 
=  C_\mathrm{I} \left(\log \frac{R_\star}{M_\star} \right)^2 +
   D_\mathrm{I}.
\end{equation}
The values of $C_\mu$, $D_\mu$, $C_\mathrm{I}$, and $D_\mathrm{I}$ for 
these fits to the spherical models are given in Table~\ref{tab:t1} for 
the five wavebands $B,V,R,I$, and $H$.  We include fits for models with 
effective temperatures $T_\mathrm{eff} = 5000$~K and models within the 
effective temperature range of $4100 - 4300$~K 
($T_\mathrm{eff} = 4200 \pm 100$~K).  Equation~(\ref{eq:mu1_rm}) and 
Eq.~(\ref{eq:imu})  will be employed in later sections.

\begin{table*}
\caption{Best-fit coefficients for Eq.~(\ref{eq:mu1_rm}) and 
         Eq.~(\ref{eq:imu}) as a function of $\log(R_\star/M_\star)$ for
         various wavebands and effective temperatures.
         \label{tab:t1} }
\begin{center}
\begin{tabular}{clll}
\hline
\hline
\multicolumn{4}{c}{$3000~\mathrm{K} < T_\mathrm{eff} \leq 8000~\mathrm{K}$} \\
\hline
Band & \multicolumn{1}{c}{$C_\mu$} & 
                         \multicolumn{1}{c}{$D_\mu$} & 
                         \multicolumn{1}{c}{$\sigma_\mu$} \\
\hline
$B$ & $2.5793 \times 10^{-4} \pm 1.537 \times 10^{-6}$ 
    & $0.738632 \pm 3.163 \times 10^{-6}$ 
    & $1.103 \times 10^{-4}$ \\
$V$ & $2.9307 \times 10^{-4} \pm 2.353 \times 10^{-6}$
    & $0.738413 \pm 4.841 \times 10^{-6}$ 
    & $1.689 \times 10^{-4}$ \\
$R$ & $3.1313 \times 10^{-4} \pm 2.745 \times 10^{-6}$ 
    & $0.738231 \pm 5.648 \times 10^{-6}$ 
    & $1.971 \times 10^{-4}$ \\
$I$ & $3.4892 \times 10^{-4} \pm 2.541 \times 10^{-6}$
    & $0.738037 \pm 5.229 \times 10^{-6}$ 
    & $1.718 \times 10^{-4}$ \\
$H$ & $4.4499 \times 10^{-4} \pm 1.775 \times 10^{-6}$ 
    & $0.737795 \pm 3.652 \times 10^{-6}$ 
    & $1.274 \times 10^{-4}$ \\
\hline
Band & \multicolumn{1}{c}{$C_\mathrm{I}$} & 
                         \multicolumn{1}{c}{$D_\mathrm{I}$} & 
                         \multicolumn{1}{c}{$\sigma_\mathrm{I}$} \\
\hline
$B$ & $0.011800 \pm 6.998 \times 10^{-5}$ 
    & $1.09614 \pm 1.440 \times 10^{-4}$ 
    & $5.024 \times 10^{-3}$ \\
$V$ & $0.013327 \pm 1.070 \times 10^{-4}$ 
    & $1.08622 \pm 2.202 \times 10^{-4}$
    & $7.683 \times 10^{-3}$\\
$R$ & $0.014227 \pm 1.246 \times 10^{-4}$
    & $1.07796 \pm 2.563 \times 10^{-4}$
    & $8.945 \times 10^{-3}$ \\
$I$ & $0.015841 \pm 1.151 \times 10^{-4}$
    & $1.06914 \pm 2.369 \times 10^{-4}$
    & $8.268 \times 10^{-3} $\\
$H$ & $0.020205 \pm 7.967 \times 10^{-5}$
    & $1.05811 \pm 1.639 \times 10^{-4}$
    & $5.720 \times 10^{-3}$ \\
\hline
\\
\multicolumn{4}{c}{$T_\mathrm{eff} = 4200 \pm 100~\mathrm{K}$} \\
\hline
Band & \multicolumn{1}{c}{$C_\mu$} & 
                         \multicolumn{1}{c}{$D_\mu$} & 
                         \multicolumn{1}{c}{$\sigma_\mu$} \\
\hline
$B$ & $2.6004 \times 10^{-4} \pm 1.800 \times 10^{-6}$ 
    & $0.738675 \pm 4.284 \times 10^{-6}$ 
    & $3.907 \times 10^{-5}$ \\
$V$ & $3.1320 \times 10^{-4} \pm 1.325 \times 10^{-6}$
    & $0.738436 \pm 3.153 \times 10^{-6}$ 
    & $2.876 \times 10^{-5}$ \\
$R$ & $3.4835 \times 10^{-4} \pm 1.332 \times 10^{-6}$ 
    & $0.738240 \pm 3.145 \times 10^{-6}$ 
    & $2.868 \times 10^{-5}$ \\
$I$ & $3.8848 \times 10^{-4} \pm 1.255 \times 10^{-6}$
    & $0.738009 \pm 2.987 \times 10^{-6}$ 
    & $2.724 \times 10^{-5}$ \\
$H$ & $4.4270 \times 10^{-4} \pm 1.088 \times 10^{-6}$ 
    & $0.737803 \pm 2.590 \times 10^{-6}$ 
    & $2.362 \times 10^{-5}$ \\
\hline
Band & \multicolumn{1}{c}{$C_\mathrm{I}$} & 
                         \multicolumn{1}{c}{$D_\mathrm{I}$} & 
                         \multicolumn{1}{c}{$\sigma_\mathrm{I}$} \\
\hline
$B$ & $0.011834 \pm 8.423 \times 10^{-5}$ 
    & $1.09813 \pm 2.004 \times 10^{-4}$ 
    & $1.828 \times 10^{-3}$ \\
$V$ & $0.014247 \pm 6.284 \times 10^{-5}$ 
    & $1.08726 \pm 1.495 \times 10^{-4}$
    & $1.364 \times 10^{-3}$\\
$R$ & $0.015834 \pm 6.260 \times 10^{-5}$
    & $1.07837 \pm 1.490 \times 10^{-4}$
    & $1.358 \times 10^{-3}$ \\
$I$ & $0.017644 \pm 5.920 \times 10^{-5}$
    & $1.06785 \pm 1.409 \times 10^{-4}$
    & $1.285 \times 10^{-3} $\\
$H$ & $0.020110 \pm 4.964 \times 10^{-5}$
    & $1.05847 \pm 1.181 \times 10^{-4}$
    & $1.077 \times 10^{-3}$ \\
\hline
\\
\multicolumn{4}{c}{$T_\mathrm{eff} = 5000~\mathrm{K}$} \\
\hline
Band & \multicolumn{1}{c}{$C_\mu$} & 
                         \multicolumn{1}{c}{$D_\mu$} & 
                         \multicolumn{1}{c}{$\sigma_\mu$} \\
\hline
$B$ & $3.0649 \times 10^{-4} \pm 1.646 \times 10^{-6}$ 
    & $0.738567 \pm 3.444 \times 10^{-6}$ 
    & $1.717 \times 10^{-5}$ \\
$V$ & $3.6147 \times 10^{-4} \pm 1.400 \times 10^{-6}$
    & $0.738322 \pm 2.930 \times 10^{-6}$ 
    & $1.461 \times 10^{-5}$ \\
$R$ & $3.9658 \times 10^{-4} \pm 1.665 \times 10^{-6}$ 
    & $0.738146 \pm 3.485 \times 10^{-6}$ 
    & $1.737 \times 10^{-5}$ \\
$I$ & $4.3136 \times 10^{-4} \pm 2.098 \times 10^{-6}$
    & $0.738956 \pm 4.391 \times 10^{-6}$ 
    & $2.188 \times 10^{-5}$ \\
$H$ & $4.9064 \times 10^{-4} \pm 2.630 \times 10^{-6}$ 
    & $0.737752 \pm 5.503 \times 10^{-6}$ 
    & $2.743 \times 10^{-5}$ \\
\hline
Band & \multicolumn{1}{c}{$C_\mathrm{I}$} & 
                         \multicolumn{1}{c}{$D_\mathrm{I}$} & 
                         \multicolumn{1}{c}{$\sigma_\mathrm{I}$} \\
\hline
$B$ & $0.013940 \pm 7.913 \times 10^{-5}$ 
    & $1.09320 \pm 1.656 \times 10^{-4}$ 
    & $8.254 \times 10^{-4}$ \\
$V$ & $0.016429 \pm 6.687 \times 10^{-5}$ 
    & $1.08206 \pm 1.399 \times 10^{-4}$
    & $6.975 \times 10^{-4}$\\
$R$ & $0.018014 \pm 7.665 \times 10^{-5}$
    & $1.07407 \pm 1.604 \times 10^{-4}$
    & $7.995 \times 10^{-4}$ \\
$I$ & $0.019579 \pm 9.420 \times 10^{-5}$
    & $1.06545 \pm 1.971 \times 10^{-4}$
    & $9.826 \times 10^{-4} $\\
$H$ & $0.022271 \pm 1.138 \times 10^{-4}$
    & $1.05616 \pm 2.381 \times 10^{-4}$
    & $1.186 \times 10^{-3}$ \\
\hline
\end{tabular}
\end{center}
\end{table*}

\section{Iterative method}
\label{sec:iter-method}

\begin{figure}[t!]
\begin{center}
  \resizebox{\hsize}{!}{\includegraphics{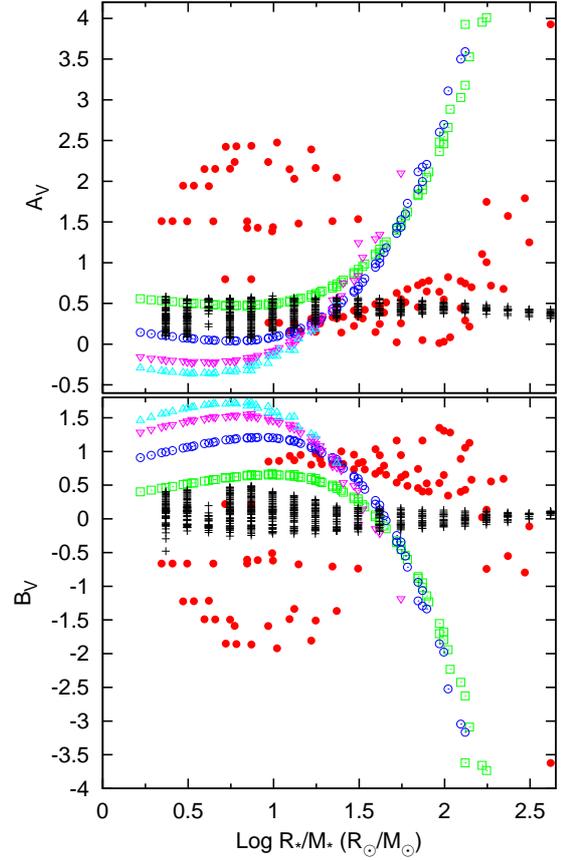}}
  \caption{$V$-band values of the A limb-darkening coefficient (top 
           panel) and the B limb-darkening coefficient (bottom panel) 
           for the linear-plus-square-root parametrization given by 
           Eq.~(\ref{eq:ldlaw}), plotted as a function of the extension 
           parameter, $R_\star/M_\star$ in solar units.  The symbols 
           have the same meaning as in Fig.~\ref{fig:v-band-logg}.}
  \label{fig:fab-ext}
\end{center}
\end{figure}

The right panels of Fig.~\ref{fig:fp-ext} show that $\mu_1$ and 
$I_\lambda(\mu_1)/2\mathcal{H}_\lambda$ depend on $R_\star/M_\star$ for the spherical 
model atmospheres.  It follows from the limb-darkening law, 
Eq.~(\ref{eq:ldlaw}), that the coefficients $A_\lambda$ and $B_\lambda$
must also be functions of the atmospheric extension, which is confirmed 
in Fig.~\ref{fig:fab-ext}.  Because the 
limb-darkening coefficients $A_\lambda$ and $B_\lambda$ in Eq.~(\ref{eq:ldlaw}) can be 
derived from observations, this establishes an observational method for 
determining the atmosphere's extension.

In Fig.~\ref{fig:fab-ext} the open colored symbols show that the 
limb-darkening coefficients of stars with $T_\mathrm{eff} > 3000$ K 
share a common dependence on the atmospheric extension for 
$\log (R_\star/M_\star) \ga 1.5$ in solar units.  This suggests it is 
possible to derive a unique value of $R_\star/M_\star$ for stars with 
atmospheric extensions $R_\star/M_\star \ga 30 \ R_{\sun}/M_{\sun}$.
Stars with $\log (R_\star/M_\star) < 1.5$ 
show a clear dependence on $T_\mathrm{eff}$.

The tight dependence of the $\mu_1$ and $I_\lambda(\mu_1)/2\mathcal{H}_\lambda$ on 
$R_\star/M_\star$ displayed in Fig.~\ref{fig:fp-ext} follows directly 
from Sect.~\ref{subsec:alpha_mu1}, where we showed that $\mu_1$ is a 
function of $\alpha_\lambda$, which depends only on $\eta_\lambda \equiv \mathcal{P}_\lambda/J_\lambda$.
The more scattered dependence of the coefficients $A_\lambda$ and $B_\lambda$ on the 
extension shown in Fig.~\ref{fig:fab-ext} is due to these coefficients 
having a less direct functional dependence on the extension through the 
separate terms $J_\lambda/2\mathcal{H}_\lambda$ and $\mathcal{P}_\lambda/2\mathcal{H}_\lambda$, as we 
showed in Paper 1. Because of this, as we will show in the following 
sections, the values of $A_\lambda$ and $B_\lambda$ do not measure atmospheric 
extension as cleanly as using $\mu_1$ and $I_\lambda(\mu_1)/2\mathcal{H}_\lambda$ as 
constraints.

Although $A_\lambda$ and $B_\lambda$ are less direct measures of the atmospheric 
extension, they are determined by the modeling of the lensing 
observations.  Therefore, we propose an iterative method to use $A_\lambda$ and 
$B_\lambda$ to determine the atmospheric extension.  The method begins by 
assuming an initial value for $\log (R_\star/M_\star)$, which is used 
to compute $\mu_1$ with Eq.~(\ref{eq:mu1_rm}).  This value of $\mu_1$ 
and the coefficients of the fit given in Table~\ref{tab:t1} enable us 
to compute the normalized intensity at the fixed point using 
Eq.~(\ref{eq:ldlaw}). Finally, we compute a new value of 
$\log (R_\star/M_\star)$ using Eq.~(\ref{eq:imu}).  This new value of 
$\log (R_\star/M_\star)$ replaces our initial assumption, starting a 
second iteration, which is continued until the value of 
$\log (R_\star/M_\star)$ converges.  If the stellar radius is 
determined by other methods, such as from interferometric observations 
and distance estimates, the converged value of 
$\log (R_\star/M_\star)$ can be solved for the star's mass.

\section{Test using a model atmosphere}
\label{sec:mod-test}

The previous section describes a potential method for determining the 
fundamental parameters of a star using stellar limb-darkening laws and 
the underlying physics behind these laws.  To test this method, 
we arbitrarily selected a model with $T_\mathrm{eff} = 5000$~K, 
$\log g = 2$, $M_\star = 5~M_{\sun}$, corresponding to 
$R_\star/M_\star = 7.42~R_{\sun}/M_{\sun}$ or  
$\log (R_\star/M_\star) = 0.87$.  Note that this value of the 
atmospheric extension falls in the region of Fig.~\ref{fig:fab-ext} 
where the limb-darkening coefficients also depend on the 
$T_\mathrm{eff}$, whereas the fixed point and intensity in 
Fig.~\ref{fig:fp-ext} are less dependent on the $T_\mathrm{eff}$.  
By selecting a trial model for which the $A_\lambda$- and $B_\lambda$-coefficients 
depend on $T_\mathrm{eff}$ as well as the atmospheric extension, we 
can determine how the iterative method is affected by this additional 
dependence.
Table~\ref{tab:t2} lists the test model's $A$ and $B$ limb-darkening 
coefficients for each of the five wavebands.
\begin{table}
\caption{Limb-darkening parameters and predicted $R_\ast/M_\ast$ for a 
         model stellar atmosphere with $T_\mathrm{eff}=5000~\mathrm{K}$,
         $\log g=2$ and $M=5~M_{\sun}$. 
         \label{tab:t2} }
\begin{center}
\begin{tabular}{crccc}
\hline
\hline
Band & \multicolumn{1}{c}{$A$} & $B$ & $R_\star/M_\star$ & $\log g$ \\
\hline
$B$ & $ 0.613$ & $0.507$ & $6.65$ & $2.09$ \\
$V$ & $ 0.048$ & $1.206$ & $6.46$ & $2.12$ \\
$R$ & $-0.300$ & $1.622$ & $7.10$ & $2.04$ \\
$I$ & $-0.642$ & $2.051$ & $7.28$ & $2.02$ \\
$H$ & $-1.189$ & $2.719$ & $7.54$ & $1.99$ \\
\hline
\end{tabular}
\end{center}
\end{table}
Using the iterative method outlined in 
Sect.~\ref{sec:iter-method}, we computed the $R_\star/M_\star$ for each 
waveband, with the results given in Table~\ref{tab:t2}.  By assuming 
a stellar radius of $R_\star = 37.1~R_{\sun}$, we determined the 
surface gravity from $R_\star/M_\star$, which is also given in 
Table~\ref{tab:t2}.

Even though the values of $A_\lambda$ and $B_\lambda$ for our test model
have no observational uncertainty, 
the results in Table~\ref{tab:t2} differ slightly from 
the test model's values.  One reason for this difference is that the 
coefficients given in Table~\ref{tab:t1} have uncertainties because they
are fits to a grid of thousands of models with the range of fundamental 
parameters given earlier; this leads to the spread of points shown in 
Fig.~\ref{fig:fp-ext}.

The variation with wavelength also contributes to the uncertainty.  
Averaging over all wavelengths, the mean difference of the predicted 
$R_\star/M_\star$ from the model's value is only $5\%$, with the 
largest deviation being about 13\% for the $V$-band.  The wavelength
variation of $R_\star/M_\star$ is related to the definition of the 
stellar radius, which we define as 
$R_\star = R(\tau_\mathrm{Ross} = 2/3)$.  However, in 
Table~\ref{tab:t2} the different wavelength bands produce values of 
$R_\star = R(\tau_\mathrm{B,V,R,I,H} = 2/3)$, which will differ slightly
from the radius where $\tau_\mathrm{Ross} = 2/3$.

Using the five wavebands in Table~\ref{tab:t2}, we find 
$(R_\star/M_\star)_\mathrm{avg} = 7.01 \pm 0.45~R_{\sun}/M_{\sun}$, and 
$\log g_\mathrm{avg} = 2.05 \pm 0.05$,
both of which are consistent with the actual 
values.  Therefore, our proposed method for using 
limb-darkening coefficients recovers the actual stellar parameters 
for the theoretical test case with a statistical uncertainty of 
$\approx 8\%$ in $R_\star/M_\star$.

We improved the test by fitting simultaneously the limb-darkening 
coefficients for all the wavelength bands using the fix-point relations 
for all models with $T_\mathrm{eff} > 3000$ K.  This resulted in the 
best-fit values $R_\star/M_\star = 6.83 \pm 0.38~R_{\sun}/M_{\sun}$, 
and $\log g = 2.04^{+0.02}_{-0.03}$, where the error bars come from the 
formal uncertainty of the fits to the coefficients in 
Table~\ref{tab:t1}, which are used in Eq.~(\ref{eq:mu1_rm}) and 
Eq.~(\ref{eq:imu}).  When the fit was repeated using only the relations for models 
with $T_\mathrm{eff} = 5000$~K, the result was 
$R_\star/M_\star = 7.61^{+0.22}_{-0.38}~R_{\sun}/M_{\sun}$ and 
$\log g = 1.99^{+0.02}_{-0.01}$.  We conclude from this test 
using a model stellar atmosphere that our proposed method for measuring 
stellar properties yields consistent results, although, as expected, 
the uncertainty of $R_\star/M_\star$ increases for stars with smaller 
atmospheric extension.
 
\section{Conclusions}

The purpose of this study has been to investigate how limb darkening 
probes fundamental stellar properties.  The limb darkening was 
parameterized by a flux-conserving linear-plus-square-root law that has 
two free parameters that are functions of two angular moments of the 
intensity.  The ratio of these two moments provides a measure of the 
amount of stellar atmospheric extension, represented by 
$R_\star/M_\star$.  We tested our method for determining the extension 
from limb-darkening parameters using a spherically symmetric model 
atmosphere and found agreement of the mean value of the five spectral 
bands analyzed to within 8\%.  

We also find significant variation of $R_\star/M_\star$  due to the definition of the stellar radius.
From the context of the model of the stellar atmosphere, we find it 
natural to define the stellar radius where $\tau_\mathrm{Ross} = 2/3$, 
but other options are possible, such as $\tau_\mathrm{Ross} = 1$ or 
using a specific wavelength, such as 500~nm, to define the radius 
where $\tau_{500} = 1$.  Observations, of course, are done in specific 
wavebands, such as $V$, and the stellar radius is assumed to refer to 
some particular optical depth, such as $\tau_\mathrm{V} = 2/3$.
Switching to another waveband will lead to another value of the radius. 
As an example, in the near-infrared the predicted radius is larger than 
at optical wavelengths.  Therefore, the definition of the stellar 
radius is ambiguous in general for studies of angular diameters 
\citep[e.g.][]{Wittkowski2004, Wittkowski2006a, Wittkowski2006b}, which 
makes the definition of the stellar radius challenging for stars with 
significant atmospheric extension.  A recent example is 
\citet{Ohnaka2011}, who measured the $K$-band angular diameter of 
Betelgeuse to be about 2 mas ($\approx$ 5\%) smaller than measured by 
\citet{Haubois2009} in the $H$-band.  However, simultaneously fitting 
$R_\star/M_\star$ to multiband limb-darkening observations provides a 
bolometric value for $R_\star/M_\star$ analogous to how spectrophotometry 
can be used to measure a star's effective temperature.

\begin{figure*}[t]
  \centering
  \resizebox{\hsize}{!}{\includegraphics{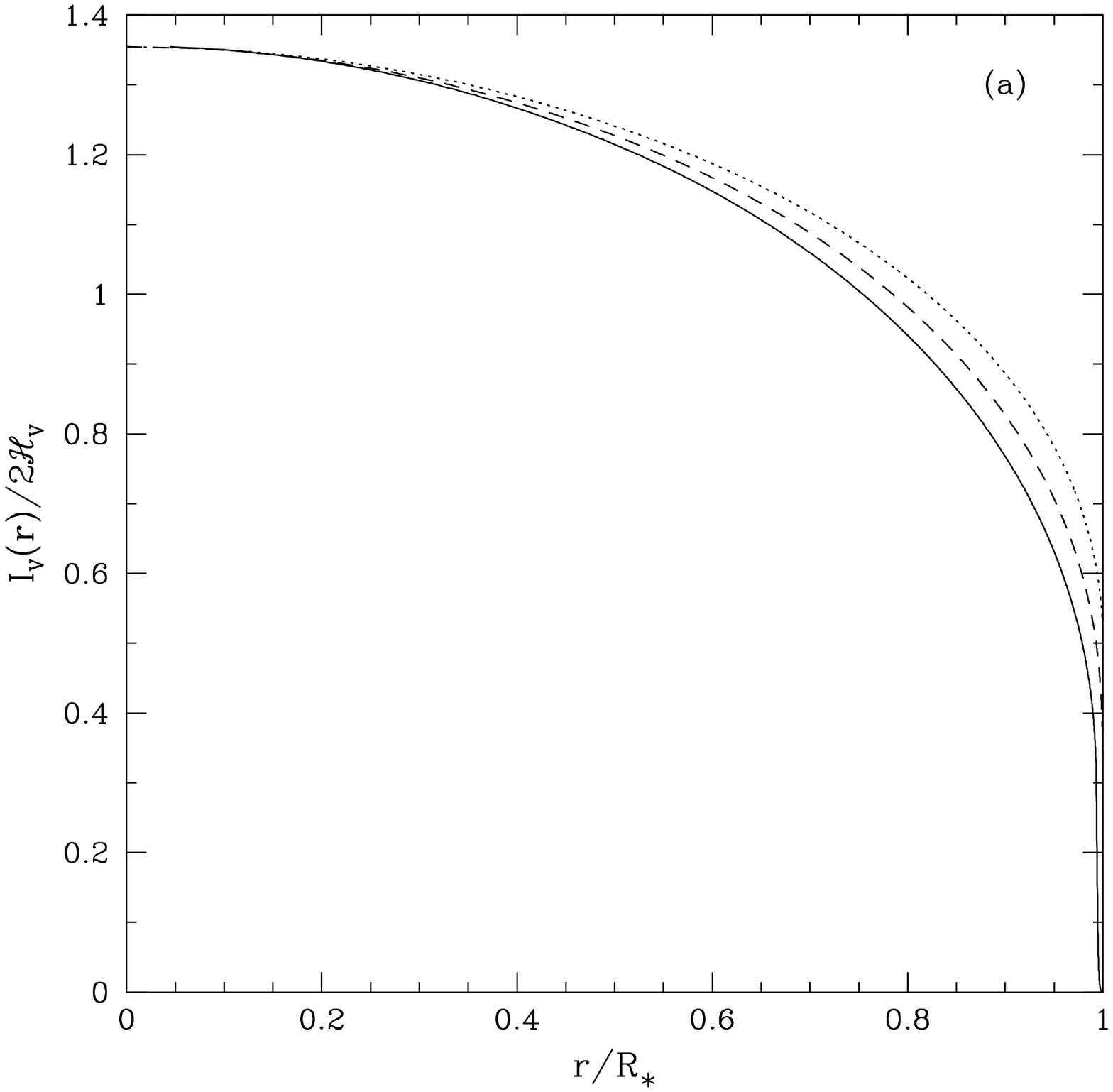}
                        \includegraphics{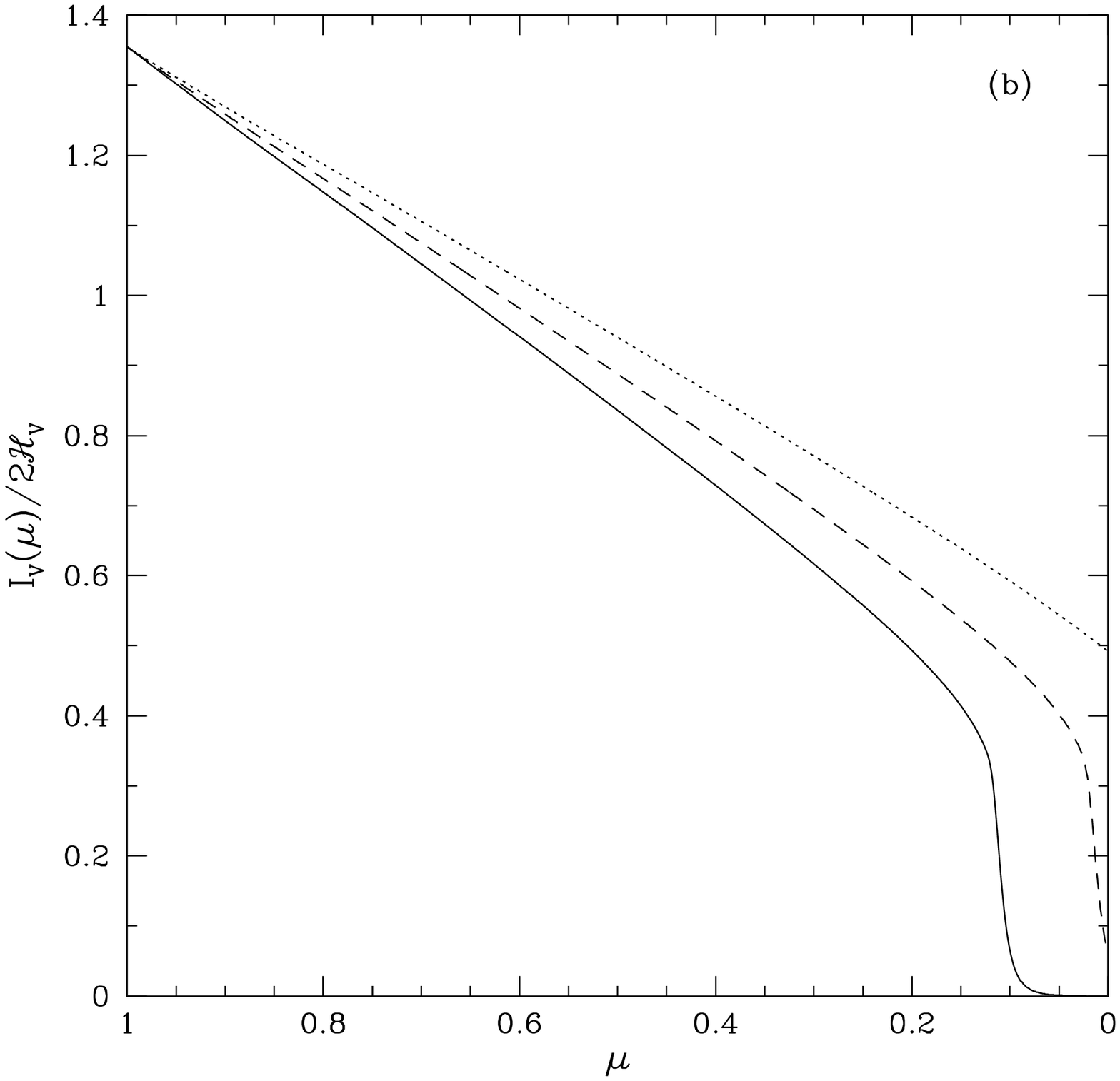}
                        \includegraphics{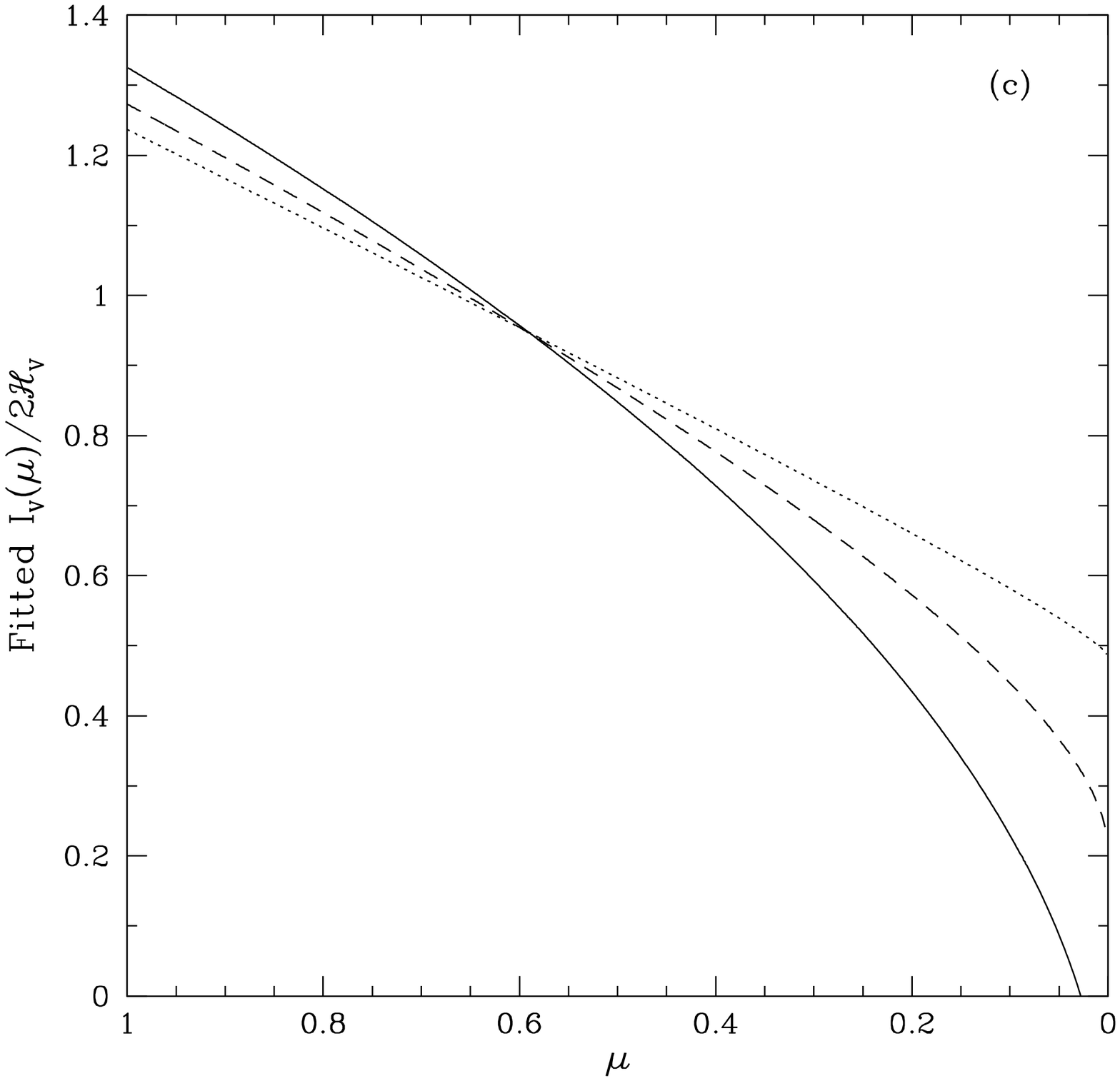}}
  \caption{(a) Surface-brightness distribution as a function of 
               fractional radius.  The solid line is the full set of 
               intensities from a spherical model atmosphere with 
               $T_\mathrm{eff} = 5000$ K, $\log g = 2.0$ and 
               $M_\star = 5~M_{\sun}$.  The dashed line has removed 
               intensity values for $r/R_\star \ge 0.995$ and then 
               renormalized the fractional radius to the interval 
               $0-1$.  The dotted line has removed intensity values 
               for $r/R_\star \ge 0.98$ and then renormalized the 
               fractional radius to the interval $0-1$.
           (b) Surface-brightness distributions from panel~(a) plotted 
               as a function of $\mu$, with the lines having the same 
               meaning as in the (a)~panel.
           (c) Fits to the surface-brightness distributions in  
               panel~(b) using Eq.~(\ref{eq:ldlaw}).  The lines have 
               the same meaning as the other two panels.}
  \label{fig:clipped}
\end{figure*}

The connection between atmospheric extension and stellar 
limb-darkening is a consequence of the assumption of spherical symmetry 
for the stellar atmospheres, not a unique feature of our \textsc{SAtlas}
code. The results presented here should also be found using intensity 
profiles from spherically-symmetric \textsc{Marcs} models and the 
\textsc{Phoenix} model atmospheres used by \citet{Claret2003} and 
\citet{Fields2003}.  Unfortunately, in both of these works the authors 
truncated the intensity profiles to remove the low intensity limb and 
then redistributed the spacing of $\mu$-points.  Truncating the 
intensity profile eliminates critical information about the limb 
darkening, and makes the star's intensity distribution resemble a 
plane-parallel atmosphere,   as we demonstrate in 
Fig.~\ref{fig:clipped}.

Our method for determining fundamental stellar parameters is based on 
the general atmospheric properties of flux-conserving limb-darkening laws,
not the particular law used here.  For instance, 
\citet{Heyrovsky2003} and \citet{Zub2011} found a fixed point in 
flux-conserving linear limb-darkening laws, which indicates that the 
linear limb-darkening coefficient is a measure of the mean intensity and flux 
of the atmosphere.  Another example is a quadratic limb-darkening law 
where the limb-darkening coefficients would provide a measure of the 
mean intensity and second moment of the intensity, $K$.  Because these 
laws measure the correlation between various moments of the intensity 
in an atmosphere, they would also measure the atmospheric extension of 
that star.  

We conclude that the method outlined in this work is a viable way to 
use a limb-darkening law, such as the linear-plus-square-root law 
employed here, to determine the fundamental physical parameters of 
stars.  The method requires knowledge of one of the degenerate 
parameters effective temperature or luminosity.  However, the method is 
able to constrain stellar parameters using limb-darkening observations 
even in the case when those limb-darkening observations from 
microlensing or eclipsing binaries cannot test the model stellar 
atmosphere directly.  It seems as though we still have things to learn 
from fairly simple representations of limb darkening, and that, as 
observations continue to improve, this will become an even more 
powerful tool in the study of stellar astrophysics.

\acknowledgements

We thank the referee for his/her many comments that led us to improve
this work. This work 
has been supported by a research grant from the Natural Sciences and 
Engineering Research Council of Canada, and HN acknowledges funding 
from the Alexander von Humboldt Foundation.  

\bibliographystyle{aa} 

\bibliography{ld_p2} 

\end{document}